\documentclass[apjl, twocolumn]{openjournal}
\usepackage{amsmath}
\usepackage{color}
\usepackage{xcolor}
\usepackage{xfrac}
\usepackage[caption=false]{subfig}
\usepackage{rotating}
\usepackage[utf8]{inputenc}
\usepackage{url}
\usepackage{listings}
\usepackage{orcidlink}
\usepackage{hyperref} 
\hypersetup{colorlinks=true,linkcolor=blue,citecolor=blue,filecolor=blue,urlcolor=blue}

\begin{document}

\shorttitle{Visualizing Disequilibrium}
\shortauthors{A. Hinkel  \& J. Vanzant}

\title{Visualizing Disequilibrium:  Tracing Two-Point Correlations to Specific Spatial Structures in the Milky Way\vspace{-14mm}}

\author{Austin Hinkel\orcidlink{0000-0002-9785-914X}$^{1\,*}$}
\author{Josephine Vanzant$^{1}$}

\affiliation{$^{1}$Department of Mathematics and Physics, Thomas More University, Crestview Hills, KY 41017}

\thanks{$^*$E-mail: \href{mailto:hinkela@thomasmore.edu}{hinkela@thomasmore.edu}}

\begin{abstract}
    The Two-Point Correlation Function can be used to demonstrate the existence of substructure within a dataset, but typical implementations do not track the individual points which contribute to each specific substructure.  In this paper, we develop an algorithm that traces a two-point correlation signal back to the contributions of each data point, allowing for structure to be visualized in novel ways.  Using this tool, we find substantial evidence for pervasive spatial clustering in the Milky Way and reveal complex undulations that warp the previously discovered vertical density waves.  We also explore the ``clumping" of stars in the thick disk and near-halo regions and discuss the possibility of dating the structure. 
\end{abstract}

\section{Introduction} \label{sec:intro}

Modern astrophysical surveys like the Gaia Space Mission \citep{prusti2016gaia, vallenari2023gaia} have produced an unprecedented amount of data, enabling the discovery and refined understanding of a number of Galactic structures and phenomena \citep{hunt2025milky}. For example, the Milky Way exhibits multiple different ``snail-like" patterns in $z - v_z$ space \citep{antoja2018dynamically, antoja2023phase, alinder2024limitations}, a warped disk \citep{kerr1957magellanic, skowron2019three}, corrugated waves \citep{minchev2009milky, xu2015rings, poggio2024great}, vertical density waves \citep{widrow2012galactoseismology, yanny2013stellar, bennett2018vertical}, stellar streams \citep{koppelman2018one}, breathing modes \citep{williams2013wobbly, asano2024growing}, and even gaseous waves like the Radcliffe Wave \citep{alves2020galactic}.  New and future surveys like the Roman Space Telescope \citep{spergel2015wide} and Vera C. Rubin Observatory \citep{ivezic2019lsst} promise to add to this list.  While such a growing compendium of structures is fascinating, it is clear that the Galaxy is complicated enough to necessitate the disentangling of various effects to better understand their origins.  
Indeed, \citet{hinkel2023two} have shown the existence of
excess, symmetry-breaking structure in the Milky Way at a number of distinct spatial scales without assuming a particular model, but could not constrain the precise structure (or structures) that contributed to that excess structure. 
To that end we develop here a tool capable of highlighting particular structures based on their spatial scales.

Fundamental to this effort is the use of the Two-Point Correlation Function (2PCF), which can identify the existence of structure at specific length scales \citep[e.g.]{wall2012practical}. Indeed, the 2PCF has been employed as a sensitive probe of structure at multiple astrophysical scales \citep[e.g.]{hauser1973statistical, lancaster2019quantifyingsmoothness, kamdar2021spatial, zhang2024using}, including within the Milky Way itself as a probe of disequilibrium effects \citep{hinkel2023two}.  While sensitive to the existence of structure beyond an assumed model, the 2PCF is limited in two important ways.  First, the 2PCF is often model-dependent, requiring comparison to a chosen model \citep[e.g.]{wall2012practical}.  Second, the two-body nature of the 2PCF necessarily ``integrates out" any star-by-star specifics.  That is, by histogramming pair-wise distances, a 2PCF analysis loses information on which stars are contributing to the correlations.

In an effort to address these limitations of a traditional 2PCF analysis and inspired by previous work on stochastic geometry and Poisson Point Processes \citep[e.g.]{chiu2013stochastic}, we develop here a novel and  powerful set of techniques to uncover and {\it visualize} the individual structures that contribute to a signal in the 2PCF.  The Two-Point Correlation Clustering algorithm (or 2PCC for short -- pronounced ``Two Pick" like the mining tool) bridges the gap between one-body and two-body analyses of structure, 
allowing one to identify specific stars which contribute to correlations in a 2PCF analysis.  The approach also allows for the incorporation of known structures to probe still finer details of the Milky Way.

The development of 2PCC is motivated by a desire to understand the nature of previous hints of excess structure in the Milky Way, particularly as it may relate to dark matter or the Galaxy's evolution. Indeed, the Two-Point Correlation studies of \citet{hinkel2023two} have employed a version of the Landy-Szalay Estimator of the 2PCF \citep{landy1993bias} to show that excess structure {\it exists} in 
volumes high above the mid-plane.
However, there has heretofore been no easy way to point to the 
particular structures in 3-dimensional position space.  That is, if structure is found to exist at some range of spatial scales, $s_1$ to $s_2$, the 2PCF does not help us visualize or localize it. The 2PCC algorithm, on the other hand, has been designed to address this limitation while allowing one to explore population statistics for stars that contribute to excess structure.
For example, do stars contributing to excess structure tend to be redder or bluer?  Do their motions differ?  Do they have different metallicities?  Might they help us date recent perturbations to the Milky Way disk?  With 2PCC, identifying individual stars contributing to structures of interest is now a possibility without full 6-Dimensional phase space data.

In this paper, we outline a mathematical description underpinning the 2PCC algorithm and develop an illustrative 1-D example (Sec.~\ref{sec:theory}). Next, we select data from the Gaia Space Mission, summarize the processes involved in the 2PCC algorithm, and effect control studies to ensure results from 2PCC are robust (Sec.~\ref{sec:method}).  In Sec.~\ref{sec:analysis}, we apply the 2PCC algorithm to a nearby sample of stars, while we discuss possible interpretations of our findings in Sec.~\ref{sec:results}.

\section{Theory} \label{sec:theory}

\subsection{Neighbor Counting Formalism}

The Two-Point Correlation Function (2PCF) is a function that, for a particular separation distance between two points, returns the probability of excess structure existing at that length scale with respect to some standard distribution \citep[see, e.g.,][]{peebles1980large}.  
The Landy-Szalay (LS) Estimator \citep{landy1993bias} is one particular way of estimating the 2PCF, and has been used to demonstrate excess asymmetric structure in the distribution of stars of the Milky Way \citep{hinkel2023two}.  Indeed, \citet{hinkel2023two} use ostensibly symmetric regions of the Milky Way to avoid the need for a standard model to which to compare data, and thus probe symmetry-breaking structure in the Galaxy at specific length scales.  Namely, an excess or dearth of the number of pairs of stars separated by some $q_i$ to $q_i + dq_i$ will result in a non-zero LS estimator at that particular scale:

\begin{equation}
    \langle \xi_{\rm LS}(q_i) \rangle = \frac{RR(q_i) - 2DR(q_i) + DD(q_i)}{RR(q_i)},
    \label{eq:x12_LS_MW}
\end{equation}
where $DD$ is a histogram of all star-star separations in the Northern half of the data set, $RR$ is a histogram of all star-star separation distances in the Southern half (reflected across $z = 0$), and DR is a histogram of all star-star cross-separations between a star in the North half of the data and a reflection of the Southern half of the data.

Critically, we would like to now highlight one particular range of length scales at a time and track how many times {\it each star} is counted in a pair-wise distance calculation, resulting in a distance between some scale $s_1$ and $s_2$.  We will represent the number of times a star is involved in a qualified pairing as $N_c$, which is the number of neighbors it has within some distance $s_2$ but beyond $s_1$.  Such a number will allow us to highlight the groups of stars responsible for a particular peak or trough in the LS Estimator-based studies of \citet{hinkel2023two}.  

Consider the case of a one-dimensional set of points distributed on the interval $[0, L]$ according to some probability density function, $f(x)$.  If one would like to find the number of pairs of points separated by a distance $s_1 < |x_2 - x_1| < s_2$ for a given location $x_1$, one must compute the following integral\footnote{For a simpler, illustrative calculation, see \citet{dunbar1997average}.}:
\begin{equation}
    N_{c}(x_1) = \int_{{\rm max}(0, \ x_1 - s_2)}^{x_1 - s_1} f(x_2) dx_2 + \int_{x_1 + s_1}^{{\rm min}(x_1 + s_2, \  L)} f(x_2) dx_2,
    \label{Eq:integralStart}
\end{equation}
where $N_{c}(x_1)$ is the number of neighbors satisfying $s_1 < |x_2 - x_1| < s_2$ for a point at $x_1$.

\subsection{Illustrative Example with a Numerical Simulation}

\begin{figure*}[]
  \begin{center}
    \subfloat[]{\includegraphics[scale=0.67]{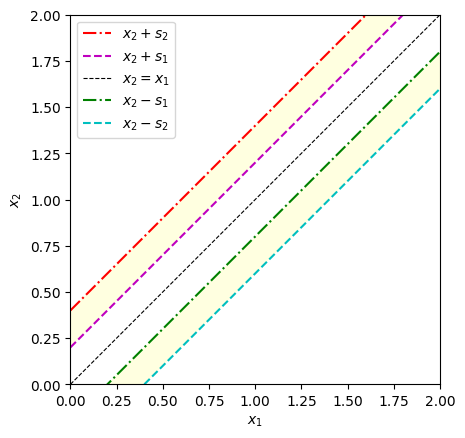}}
    \subfloat[]{\includegraphics[scale=0.67]{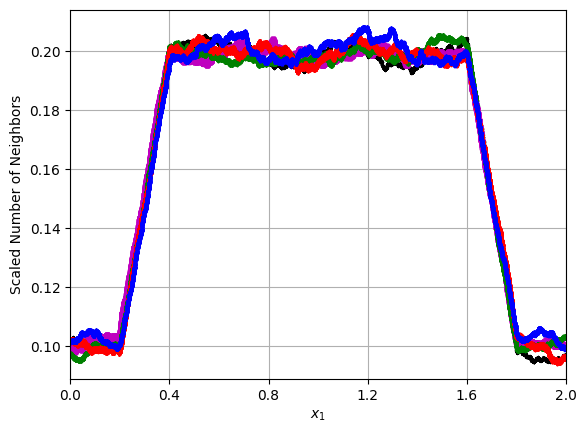}}
    \caption{
    (a) A representation of all possible pairings of points, $(x_1, x_2)$, on the interval $[0, L]$, for $L = 2$. Valid pairings are shaded in a light yellow color, where we have chosen $s_1 = 0.2$ and $s_2 = 0.4$ for this illustration.  
    (b) The results of counting the number of neighbors for points at a given $x_1$ from 5 simulations of $N = 20,000$ points on the interval $[0, L]$, for $L = 2$, $s_1 = 0.2$, and $s_2 = 0.4$.  The number of neighbors have been divided by the number of total points so that $N_c$ is represented by a number between 0 and 1.
    }
  \label{fig:RegionsAndSim}
  \end{center}
\end{figure*}

For the simplest case of a uniform, one-dimensional distribution of points on the interval $[0, L]$, $f(x) = \sfrac{1}{L}$ and Eq.~\ref{Eq:integralStart} can be solved exactly.  That is,
\begin{equation}
    N_{c}(x_1) = \frac{1}{L} \int_{{\rm max}(0, \ x_1 - s_2)}^{x_1 - s_1} dx_2 + \frac{1}{L} \int_{x_1 + s_1}^{{\rm min}(x_1 + s_2, \  L)} dx_2,
    \label{Eq:integralSpecialCase}
\end{equation}
which results in the following piecewise result for $N_{c}(x_1)$:
\begin{equation}
    N_{c}(x_1) = \frac{1}{L} \begin{cases}
    s_2 - s_1 & 0 < x_1 \leq s1 \\
    x_1 + s_2 - 2s_1 & s_1 < x_1 \leq s_2 \\
    2(s_2 - s_1) & s_2 < x_1 \leq L - s_2 \\
    L - x_1 + s_2 - 2s_1 & L - s_2 < x_1 \leq L - s_1 \\
    s_2 - s_1 & L - s_1 < x_1 \leq L,
    \end{cases}
    \label{Eq:piecewiseSpecialCase}
\end{equation}
where we note that the five branches of the piecewise function above are determined by the five regions of $x_1$ with unique pairs of integral bounds as depicted bounding the shaded regions in Fig.~\ref{fig:RegionsAndSim}(a).  Namely, $x_1 < s_1$ can only have neighbors above the $x_2 = x_1$ line, in the upper shaded region.  Then, from $s_1 < x_1 < s_2$, the lower shaded region begins to contribute neighbors to the count.  For $s_2 < x_1 < L - s_2$, neighbors from both shaded regions contribute to the count, where we note the lower bound of the bottom region has changed at $x_1 = s_2$.  Similarly, for $L - s_2 < x_1 < L - s_1$, the upper bound of the upper region changes, and so fewer neighbors are found for this region.  Finally, for $L - s_1 < x_1 < L$, only the neighbors in the bottom region contribute.

Choosing $L = 2$, $s_1 = 0.2$, and $s_2 = 0.4$, we find that
\begin{equation}
    N_{c}(x_1)  = \begin{cases}
    0.1 & 0.0 < x_1 \leq 0.2 \\
    \frac{x_1}{2}  & 0.2 < x_1 \leq 0.4 \\
    0.2 & 0.4 < x_1 \leq 1.6 \\
    \frac{2.0 - x_1}{2}  & 1.6 < x_1 \leq 1.8 \\
    0.1 & 1.8 < x_1 \leq 2.0.
    \end{cases}
    \label{Eq:piecewiseSimulation}
\end{equation}
These values for $N_{c}(x_1)$ are borne out in five simulations of 20,000 points each wherein points are randomly drawn on the interval $[0, 2]$, and the number of neighbors for a given point are counted at each $x_1$.  The result of these simulations is shown in Fig.~\ref{fig:RegionsAndSim}(b), where the number of neighbors has been divided by the total number of particles in the simulation ($N = 20,000$) to aid in comparisons between simulations.

As the distribution of points on $[0, L]$ is uniform, the results are rather intuitive.  Points closer to an edge than $s_1$ can only pair with points on one side, while points in the center can pair with points on either side, resulting in the factor of two difference between the central branch of the piecewise function compared to the outermost branches.  The linear transition regions arise from the points at least $s_1$ from an edge but within $s_2$ of that edge.  In this case, more and more pairings become possible as one moves towards the center.  

Although simple, this example provides a more intuitive feel for the type of neighbor counting we are performing.  In particular, note that if there were any excess structures beyond the expected uniform distribution, it would have to be interpreted in the context of the region of $x_1$ where it is found.  That is, both the background distribution of points and the boundaries of the sample are important.  
We discuss this matter further in the methodology section.

\section{Methodology} \label{sec:method}

\subsection{Data Selection}

The data for this analysis are from the Gaia Data Release 3 (DR3) archive \citep{prusti2016gaia, vallenari2023gaia}, focusing on a volume similar to \citet{hinkel2023two}, although we note that they use Data Release 2 data.  To start, we use Gaia DR3 stars from the \texttt{gaia\_source\_lite} table with five parameter astrometric solutions that satisfy: 
$|b| > 30^{\circ}$,
$14 < G < 18$ mag,
$\varpi > 0.32$ mas,
$0.5 < G_{\rm BP} - G_{\rm RP} < 2.5$ mag, and
$(\sfrac{1}{\varpi}){\rm cos}(b) < 0.35 $ kpc, where many of these geometric cuts are made in the query itself to cull the number of stars for later, more restrictive cuts on quantities not indexed in the Gaia database.  We include them here purely for transparency.  Additionally, we note that five-parameter astrometric sources are used, as six-parameter sources are reported to be more likely to have problems associated with them \citep{lindegren2021gaia}.

Critically, these cuts follow in the manner of \citet{GHY20} and \citet{HGY20} in order to obtain a complete \citep{everall2022completeness} sample of nearby stars without the imprint of the Gaia scan law while also circumscribing the smaller, reflection-symmetric volume we wish to analyze, though we note that we do not shift the parallax measurements like \citet{GHY20}, as the stars are relatively close by, and thus we do not expect a two-body analysis to be sensitive to a small zero-point parallax offset.  We also forgo the box cuts of \citet{GHY20} and \citet{HGY20} as we do not expect the Large and Small Magellanic Clouds to contaminate our nearby sample.
In the spirit of reproducibility, please also note that the first condition ($|b| > 30^{\circ}$) was split into two queries, $b > 30^{\circ}$ and $b < -30^{\circ}$, which were performed separately. These separate queries result in 1,505,754 and 1,682,731 stars respectively, for a total of 3,188,485 stars. 

With the queried data in hand, further, more restrictive cuts are made in post-processing.  Sticking with the right-handed coordinate system employed in \cite{hinkel2023two} where the Sun lies on the negative $X$ axis and at $Z = 0$, we keep stars for which the heights above or below the Galactic midplane satisfy $0.2 < |z| < 2.0$ kpc, the Galactocentric, in-plane radial coordinates fall within $7.8 < R < 8.2$ kpc.\footnote{For simplicity, we adopt a Sun-Galactic Center distance of $R_0 = 8.0$ kpc, though we note more precise determinations exist \citep{abuter2019geometric}. We also neglect the Sun-mid-plane offset in keeping with \cite{hinkel2023two}.}, and for which Galactocentric azimuth coordinates fall within $178 < \phi < 182$ deg.  These particular selections are chosen to coincide with the analysis in Figure 6 of \citet{hinkel2023two}, which motivated this work, and intentionally exclude cuts on relative parallax error, as motivated in later sections.  After all of these cuts are made, 1,534,449 stars remain.

\subsection{The Algorithm} \label{subsec:algo} 

The Two-Point Correlation Clustering algorithm, or 2PCC, extracts the number of times each point in a data set is involved in a valid pairwise distance, $s$, that falls between $s_1 < s < s_2$.  In effect, this highlights spatial structure at specific length scales between some lower limit $s_1$ and upper limit $s_2$.  To accomplish this, 
the stars in our analysis are split into two, three-dimensional \texttt{KDTree}s using the scikitlearn library \citep{scikit-learn}. These two binary trees separate stars in the North and stars in the South for detection of symmetry-breaking effects.

For each of the two trees, the number of neighbor stars within some distance $s_2$ is calculated for each star.  We denote this number as $N_{c}^{s_2}(\vec{r}_i)$, although we shorten this somewhat to $N_{c_i}^{s_2}$ for notational compactness.  The number of neighboring stars is again calculated for some smaller separation scale $s_1$, which we call $N_{c_i}^{s_1}$.  The difference in the number of neighbors at each scale is then $N_{c_i} = N_{c_i}^{s_2} - N_{c_i}^{s_1}$, so that for some choice of $s_1$ and $s_2$, every star has a corresponding $N_{c_i}$ that reveals the number of times it is involved in a pairing of stars with a separation distance between $s_1$ and $s_2$.  

Further, to add additional context to the findings of \citet{hinkel2023two}, we choose to study symmetry-breaking effects in the values of $N_{c_i}$, as opposed to comparing the data against a model galaxy as is typical of 2PCF studies.  
In practice, this involves the following process, which is repeated for each volume studied:

\begin{itemize}
    \item For a particular choice of $s_1$ and $s_2$, $N_{c_i}$ is calculated separately for the North and South halves of the data set.  This results in a value of $N_{c_i}$ for each star.  We note that we neglect any Sun-midplane offset.
    \item The stars in the South half of the data ($Z < 0$) are then reflected across the Galactic midplane ($Z \rightarrow |Z|$) to aid in comparison with the Northern set of stars.
    \item In order to compare the two hemispheres, a k-Nearest-Neighbors Regression model is trained on the southern half of the data, where the Galactocentric Rectangular Coordinates, $(X, Y, Z)$, are used as input features and $N_{c_i}$ as the target variable. 
    \item The k-Nearest-Neighbor Regressor is then tasked with predicting the ``expected pairings," $N_{\rm exp_i}$, for stars in the North based on the training data from the South.  To do this, the values of $N_{c_i}$ for the five closest neighbors from the reflected South data are averaged.
    \item Finally, we define the excess neighbor ratio, $r_i = \left( \sfrac{N_{c_i}}{ N_{\rm exp_i}} \right) \cdot \left( \sfrac{N_S}{N_N} \right)$, where we multiply by the ratio of neighbors to expected neighbors by the number of stars in the south, $N_S$ divided by the number of stars in the North, $N_N$ in order to account for the difference in the number of stars North and South of the midplane.  Therefore, $r_i > 1$ implies an excess of neighbors in the North, while $r_i < 1$ implies an excess of neighbors in the South (or dearth in the North).  The absence of symmetry-breaking substructure will result in $r_i = 1$.  
\end{itemize}
Because the volumes selected in the North and in the South are reflection-symmetric, we eliminate the boundary effects discussed at the end of Sec.~\ref{sec:theory} by constructing the excess neighbor ratio, $r_i$. 

Upon completion of the algorithm outlined above, all stars in the North have an associated value of $r_i$.  The value of $r_i$ is then passed as the color argument for a scatter plot of the Northern stars to visualize symmetry-breaking structure.  A similar process can be used for the Southern stars.

\vspace{\medskipamount}

\subsection{Uncertainty in the Excess Neighbor Ratio}

To estimate the uncertainty of our assessment of the excess ratio, we make a simplifying assumption that the stars are uncorrelated such that the number of neighbors within some distance of a given star follow Poissonian statistics.  That is, for a star with $N_{c_i}$ neighbors, the uncertainty in the number of neighbors is $\sigma_{N_{c_i}} = \sqrt{N_{c_i}}$.  We make a similar simplification in assessing the uncertainty in the expected neighbors, $\sigma_{N_{\rm exp_i}} = \sqrt{N_{\rm exp_i}}$.  Finally, assuming zero uncertainty in the total number of stars in the North ($N_N$) and South ($N_S$), we then approximate the uncertainty in the assessment of $r_i$ in the usual way as: 

\begin{equation}
    \sigma_{r} = \sqrt{\left(\frac{\partial r}{\partial N_{c}}\right)^2 \sigma_{N_{c}}^2 + \left(\frac{\partial r}{\partial N_{\rm exp}}\right)^2 \sigma_{N_{\rm exp}}^2}
    \label{Eq:sigma_r_init}
\end{equation}

which simplifies to:

\begin{equation}
    \sigma_{r_i} = r_i \cdot \sqrt{\frac{N_{c_i} + N_{\rm exp_i}}{N_{c_i} \cdot N_{\rm exp_i}}}.
    \label{Eq:sigma_r}
\end{equation}

Upon completion of the algorithm described in Subsec.~\ref{subsec:algo}, all stars in the Northern data set have an associated value of $r_i$ and a corresponding uncertainty $\sigma_{r_i}$.

\subsection{Control Studies}

\begin{figure*}[]
  \begin{center}
    \subfloat[]{\includegraphics[scale=0.82]{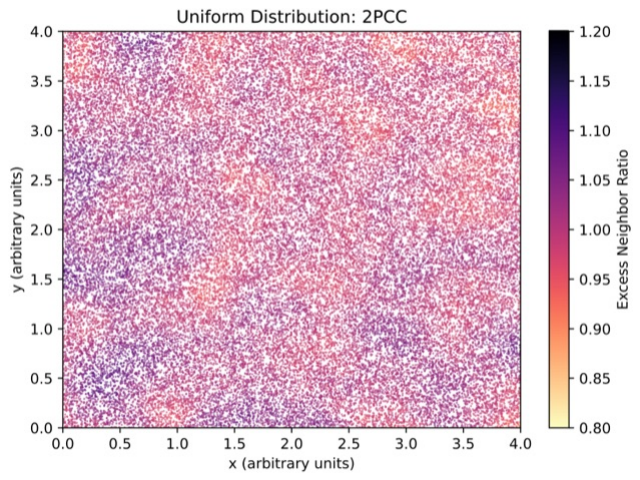}}
    \subfloat[]{\includegraphics[scale=0.82]{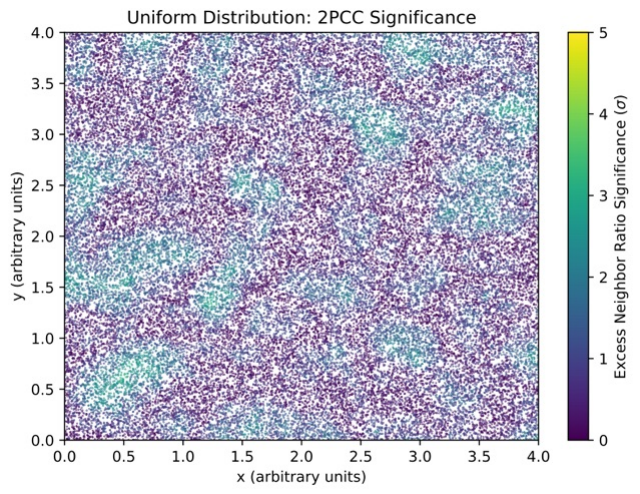}}

    \subfloat[]{\includegraphics[scale=0.57]{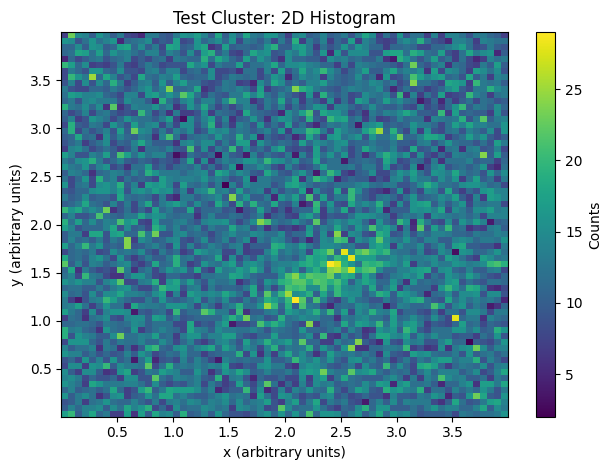}}
    \subfloat[]{\includegraphics[scale=0.82]{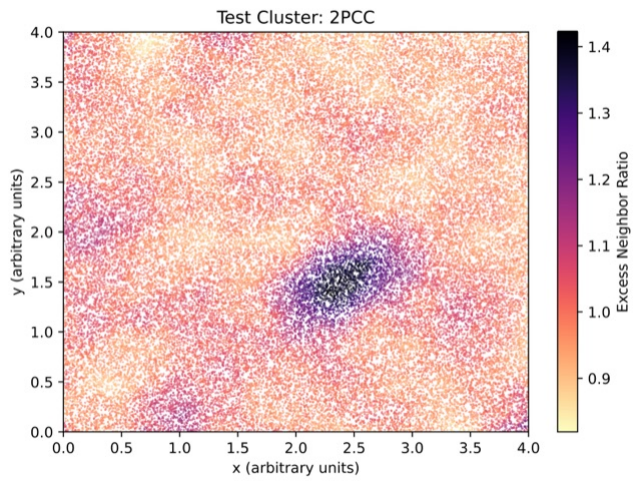}}
    \caption{
    (a) A 2PCC analysis of a uniform distribution of points returns Expected Pair Ratios near unity, as expected for structureless data.
    (b) Additionally, the points with Excess Pair Ratio $r \neq 1$ are predominantly within 2 standard deviations of $r = 1$.
    (c) Simulated structure (1,000 points) added to the uniform background of 50,000 points can be seen in a 2D histogram, but some of the detail is lost in the binning process.
    (d) In contrast, a 2PCC analysis of the same simulated structure reveals the location and shape of the cluster of point in greater detail. The clustered points are identified as $5\sigma$ or greater significance in contrast to panel b of this figure. 
    }
  \label{fig:Control}
  \end{center}
\end{figure*}

In an effort to ensure that the 2PCC algorithm does not falsely indicate structure where none exists, we demonstrate here how 2PCC responds to pseudo-randomly drawn, uniformly distributed data sets of $N = 50,000$ points. In other words, we do not expect two uniformly distributed samples to yield significant structural differences.  This is explicitly tested in panel (a) of Fig.\ref{fig:Control}, with the significance of the results plotted in panel (b).  Although the excess neighbor ratio, $r_i$, is not perfectly 1 for all points in this control study, it is clear from panel (b) that the departures from $r_i = 1$ are not significant, with only a few percent of points beyond $2\sigma$ in either direction from an $r_i$ value of unity.

As an additional illustration of the effectiveness of 2PCC, we add clustered structure to a pseudo-randomly drawn, uniform background distribution and visualize this structure via a 2-D histogram in panel (c) of Fig.\ref{fig:Control}.  Reanalysis of the same data with 2PCC in panel (d) emphasizes the structure more effectively against the background and avoids binning effects.  Still, the two methods agree on the location and rough shape of the simulated cluster.


\subsection{Testing the k-Nearest Neighbors Model}

Finally, we verify the performance of the k-Nearest-Neighbor Regressor that was trained on the Southern half of the data.  Specifically, after determining the number of neighbors for each star in the South, we split the data into a training set (80\%) and test set (20\%), and train the regressor on the former.  The model is then tasked with predicting the number of neighbors for the held-out test set before the model is deployed to assess any North/South asymmetry. 

\begin{figure}[h]
  \begin{center}
    \includegraphics[scale=0.79]{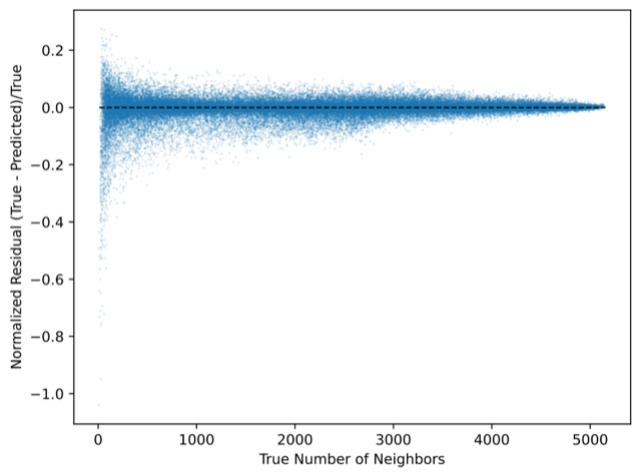}
    \caption{(a) The k-Nearest-Neighbor Regressor predicts the true number of neighbors for nearly all stars in the South, save for some stars with very few neighbors, as evidenced by normalized residuals falling very close to zero.  The small number of stars with the largest normalized absolute residuals ($>$ 10\%) are largely along the edges of the volume.  As the Excess Neighbor Ratio, $r_i$, is constructed by dividing the number of neighbors in the North by the output of the k-Nearest-Neighbor Regressor, both the numerator and denominator are affected similarly by these edge effects.  }
  \label{fig:knn}
  \end{center}
\end{figure}

Employing multiple iterations of this test/train split procedure, we arrive at a model that consistently recovers the known number of neighbors across the full range of the target variable (the normalized residuals are plotted in Fig.~\ref{fig:knn} panel a).  The predictions of the k-Nearest-Neighbor model with the largest normalized absolute residuals (in excess of 10\%) are largely along the border of the data set (i.e., those with artificially fewer neighbors due to cuts) and the residuals do not exhibit wave-like trends.  As the Northern half of the data is reflection symmetric, it, too, suffers from this same edge effect, which is then ``canceled" in the construction of $r_i$.

We also note that the k-Nearest-Neighbor hyperparameter, $k$, that minimized the root-mean-squared error was around $5 \leq k \leq 10$, with all giving comparable performances.  As larger values of $k$ tend to obscure small scale structure, we have opted to use $k = 5$ for this work.  Nonetheless, we have verified that our results are robust to small changes in the choice of $k$.

\section{Analysis} \label{sec:analysis}


\subsection{The Gaia Data}

As a proof-of-concept of our method, we examine the asymmetric vertical density waves in the solar neighborhood described by  \citet{widrow2012galactoseismology} and \citet{bennett2018vertical}, among others.  Namely, these previous studies have assessed the asymmetry of the vertical structure of the Galaxy, which has necessitated binning of the data, and thus loss of information on individual stars.  In an effort to detect the vertical density waves as a test of our method (and to better visualize them), we apply 2PCC to a similar volume of space, as detailed in the Data Selection subsection.  

\begin{figure*}[]
  \begin{center}
    \subfloat[]{\includegraphics[scale=0.82]{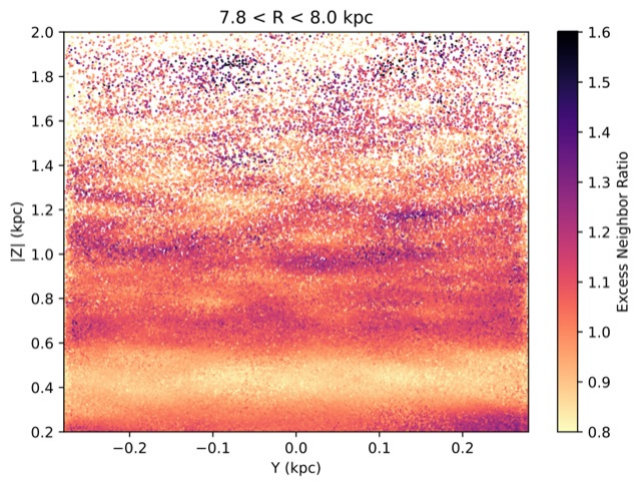}}
    \subfloat[]{\includegraphics[scale=0.82]{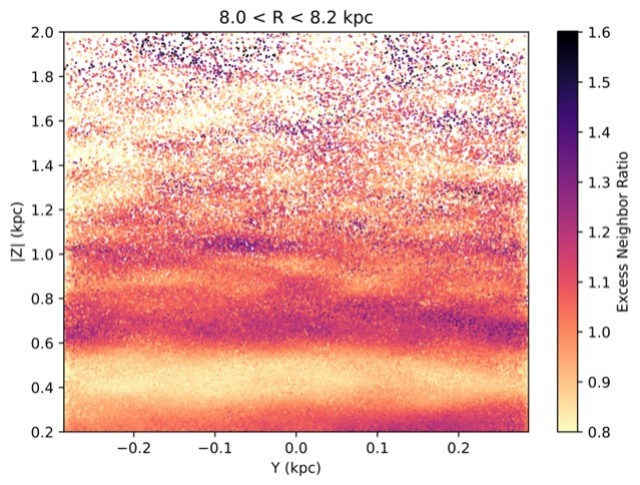}}

    \subfloat[]{\includegraphics[scale=0.82]{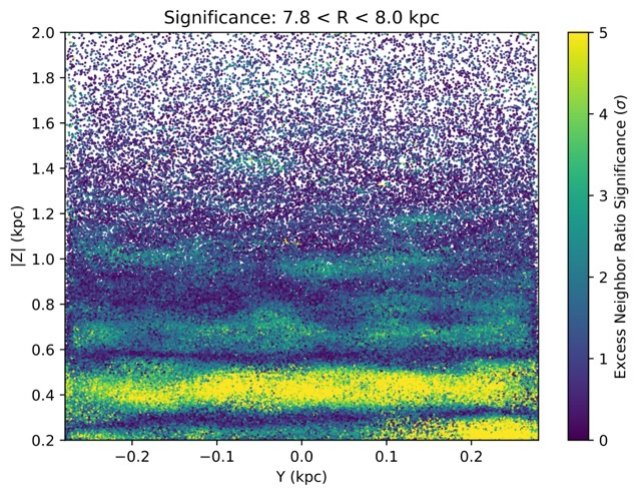}}
    \subfloat[]{\includegraphics[scale=0.82]{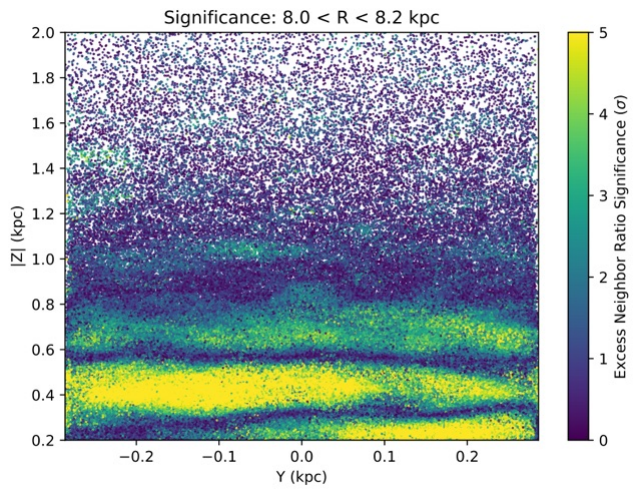}}
    \caption{
    Structure at scales from $s_1 = 0.01$ kpc to $s_2 = 0.05$ kpc, depicted in the $Y-Z$ plane, for (a) $7.8 < R < 8.0$ kpc and (b) $8.0 < R < 8.2$ kpc.  All other cuts on the data are outlined in Sec.~\ref{sec:method}.  The vertical waves discovered by \citet{widrow2012galactoseismology} are apparent in both panel a and b, with a clear indication that the waves are not perfectly planar, but rather exhibit radial and azimuthal differences.  There is additional vertical structure at higher $|z|$ that is less coherent than the waves near the mid-plane, though we note it does not rise to the level of significance of the lower-$|Z|$ structure for $7.8 < R < 8.0$ kpc or $8.0 < R < 8.2$ kpc, as depicted by panels (c) and (d) respectively.  
    }
  \label{fig:verticalWavesPOC}
  \end{center}
\end{figure*}

Choosing $s_1 = 0.01$ kpc and $s_2 = 0.05$ kpc (we will return to this choice later), we show that we can reproduce and {\it visualize} the vertical waves \citep[e.g.]{bennett2018vertical} for $7.8 < R < 8.0$ kpc (Fig.~\ref{fig:verticalWavesPOC}a) and $8.0 < R < 8.2$ kpc (Fig.~\ref{fig:verticalWavesPOC}b).  Namely, we see:
\begin{itemize} 
    \item a region of stars with $r_i > 1$ near $|Z| \sim 0.7$ kpc, implying a north-heavy asymmetry of stars,
    \item a region of stars with $r_i < 1$ near $|Z| \sim 0.4$ kpc, implying a south-heavy asymmetry of stars, 
    \item and a hint of another north-heavy ($r_i > 1$) peak near $|Z| \sim 0.2$ kpc, though our cuts on the data prevent us from following the peak across a range of azimuth.\footnote{Here, 
    we have adopted Galactocentric Rectangular Coordinates 
    for the visualization of the data, with $X$ increasing 
    from the Galactic Center in a direction directly away from the Sun,
    $Z$ increasing towards the North Galactic Pole, 
    and $Y$ defined to complete the right-handed coordinate system.  
    For the special case of stars very near the Sun, 
    $Y$ is a proxy for the azimuthal direction.  
    We neglect the Sun-midplane offset for simplicity.} 
\end{itemize}

These local density variations are found to be statistically significant (Fig.~\ref{fig:verticalWavesPOC} panels c and d) and coincide with the peaks in Figure 5 of \citet{bennett2018vertical}, indicating that we are successfully tracing the asymmetry back to the individual stars that participate in the effect.

Moreover, we see that the coherent, vertical structure varies slightly with the selected radial bins (c.f. panels a and b of Fig.~\ref{fig:verticalWavesPOC}, particularly for $|Z| \in [0.7, 1.3]$ kpc), and substantially differs at higher values of $|Z|$.  In particular, azimuthal variations appear along the vertical waves, in the form of a gradual bending and smaller scale bifurcations and undulations, likely related to the findings of \citet{ferguson2017milky}.  These changes in the vertical waves are also broadly consistent with the Two-Point Correlation Function results of \citet{hinkel2023two}.  Namely, Figure 6 of \citet{hinkel2023two} indicated an excess of vertical structure on small scales across their sample with very consistent results in the low-$|Z|$ region, while also indicating a wide variety of higher-$|Z|$ vertical structure.  We have visualized some of these correlations in Fig.~\ref{fig:verticalWavesPOC}.  

\begin{figure*}[]
  \begin{center}
    \subfloat[]{\includegraphics[scale=0.82]{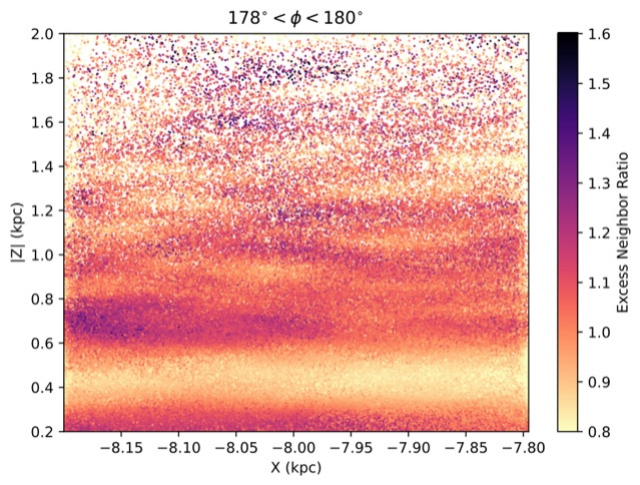}}
    \subfloat[]{\includegraphics[scale=0.82]{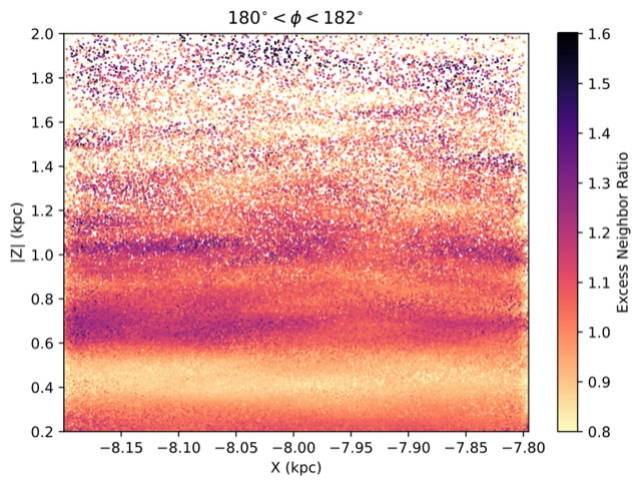}}

    \subfloat[]{\includegraphics[scale=0.82]{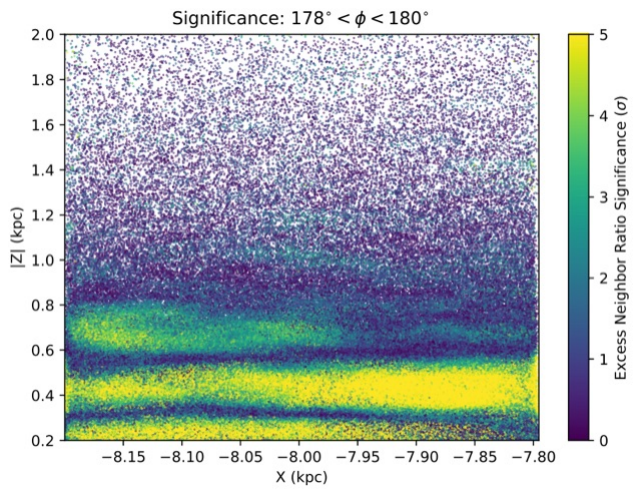}}
    \subfloat[]{\includegraphics[scale=0.82]{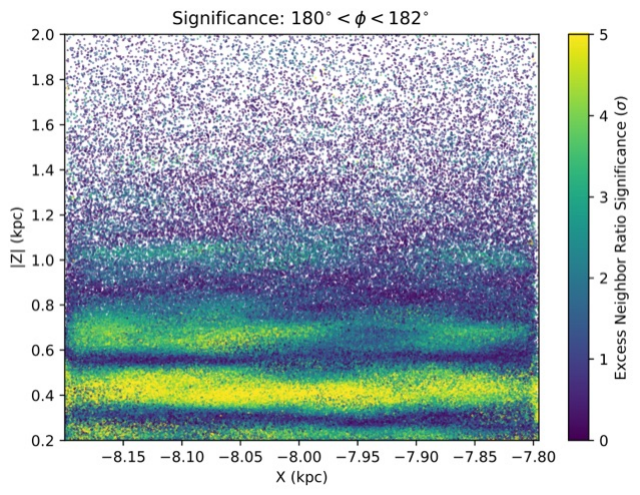}}
    \caption{
    Structure at scales from $s_1 = 0.01$ kpc to $s_2 = 0.05$ kpc, depicted in the $X-Z$ plane, for $7.8 < R < 8.2$ kpc and (a) $178 < \phi < 180$ deg, (b) $180 < \phi < 182$ deg.  All other cuts on the data are outlined in Sec.~\ref{sec:method}.  Similar to Fig.~\ref{fig:verticalWavesPOC}, the vertical waves are clear in both panels, with substantial radial and azimuthal variations.  Additionally, there is additional vertical structure at higher $|z|$ that is less coherent than the waves near the mid-plane, though we note it does not rise to the significance of the structure at lower $|Z|$ as depicted in panel (c) for the $178 < \phi < 180$ deg. data and (d) for the $180 < \phi < 182$ deg. data.
    }
  \label{fig:verticalWavesPOC_radial}
  \end{center}
\end{figure*}

We further trace the implied radial variations from the first two panels of Fig.~\ref{fig:verticalWavesPOC} in Fig.~\ref{fig:verticalWavesPOC_radial} for different azimuthal slices of the volume.  Namely, adopting the same choice of $s_1$ and $s_2$, we visualize the radial variations of the vertical waves for (a) $178 < \phi < 180$ deg. and (b) $180 < \phi < 182$ deg.  Again, the vertical waves are largely planar, but with clear radial and azimuthal differences that rise to high statistical significance (panels c and d).

\begin{figure*}[]
  \begin{center}
    \subfloat[]{\includegraphics[scale=0.82]{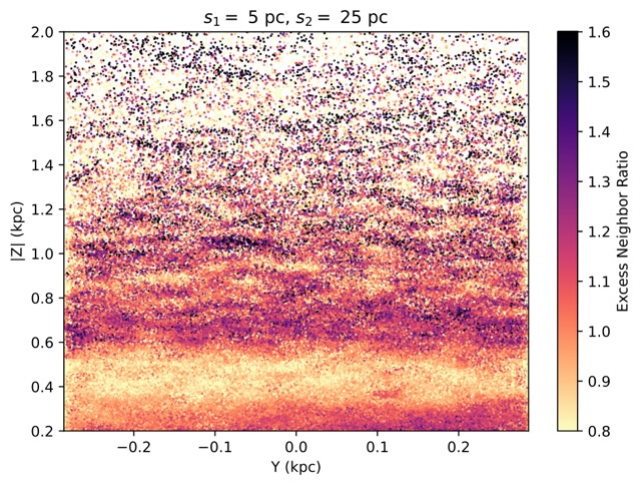}}
    \subfloat[]{\includegraphics[scale=0.82]{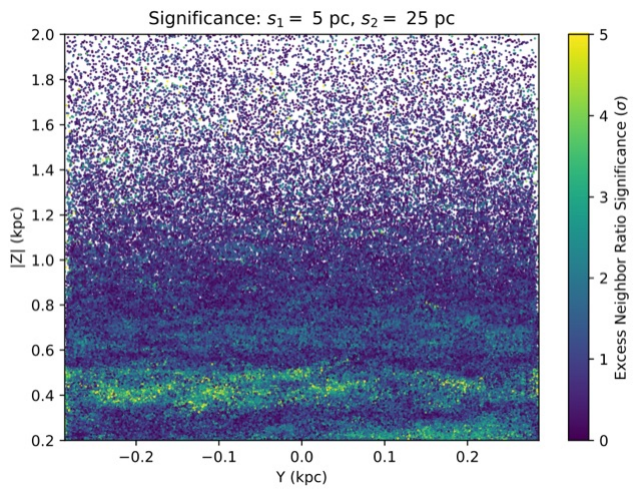}}
    
    \subfloat[]{\includegraphics[scale=0.82]{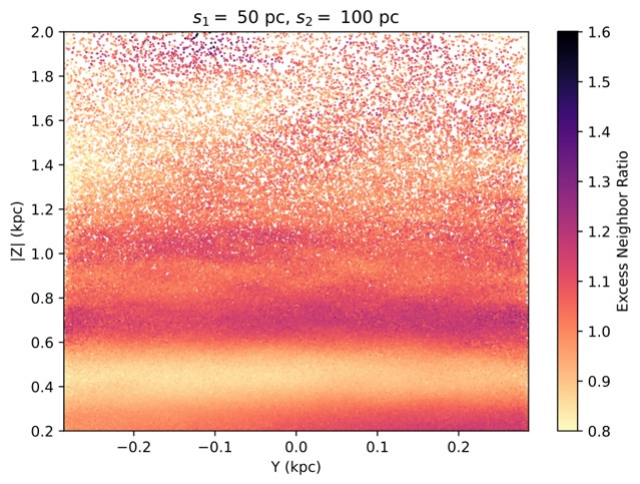}}
    \subfloat[]{\includegraphics[scale=0.82]{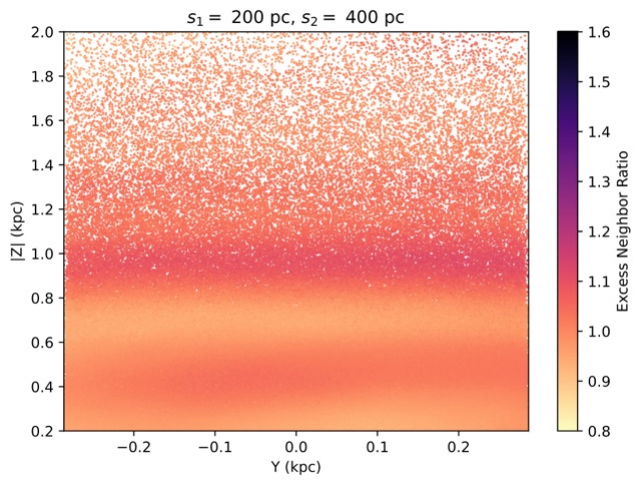}}
    \caption{
    Structure for stars with $8.0 < R < 8.2$ kpc at scales of 
    (a) $s_1 = 0.005$ kpc to $s_2 = 0.025$ kpc shows large departures from the expected $r = 1$ case for a symmetric, structureless disk, though
    (b) the deviations are largely insignificant for the high-$|Z|$ data as very few neighboring stars exist within scales of $s_2 = 0.025$ kpc.
    (c) Structure for stars with $8.0 < R < 8.2$ kpc at scales of $s_1 = 0.05$ kpc to $s_2 = 0.1$ kpc and for
    (d) $s_1 = 0.2$ kpc to $s_2 = 0.4$ kpc.
    While our ability to resolve the vertical waves decreases with larger choices of $s_1, s_2$, some differences can be spotted at high-$|z|$.  It should also be noted that the wave-like trend in the Excess Neighbor Ratio reverses at scales of $s_1 = 0.2$ kpc to $s_2 = 0.4$ kpc as this separation distance corresponds to stars with pairs including one member in between the wave crests with a multitude of neighbors falling in both adjacent wave crests.
    }
  \label{fig:yz_multipleScales8082}
  \end{center}
\end{figure*}

We have also examined the same regions of space as Fig.~\ref{fig:verticalWavesPOC} for different choices of $s_1$ and $s_2$ in order to understand the significance of the more diffuse, high-$|Z|$ structure indicated in \citet{hinkel2023two}.  Indeed, Fig.~\ref{fig:yz_multipleScales8082} shows a persistent vertical wave across a number of different choices of $s_1$ and $s_2$, but the smaller-scale choices appear to show more detail at high-$|Z|$.  We caution, however, that the choice of small scales for $s_1$ and $s_2$ results in few pairs available to each star in the sample, limiting the statistical significance of the analysis.  Indeed, Fig.~\ref{fig:yz_multipleScales8082}(a) depicts a number of clumps of stars at high-$|Z|$ with values for the Excess Neighbor Ratio in excess of $\sim 1.4$, though the statistical significance for each star's {\it individual} Excess Neighbor Ratio is shown to be largely insignificant in panel (b).  Larger choices of $s_1$ and $s_2$ (panels c and d) do not indicate substantial, high-$|Z|$ structure, but still exhibit some azimuthal heterogeneity, which is worth exploring further in light of the  tendency for distance uncertainty to increase with distance.

Clearly, there is a trade-off between zeroing in on finer Galactic structure by lowering $s_1$ and $s_2$ and running into issues with parallax uncertainties and losing statistical strength as the number of neighbors within some small distance begins to drop.  Our choice of $s_1 = 0.01$ kpc and $s_2 = 0.05$ kpc enables a clear view of the substructure at low-$|Z|$ where data are of high significance and very low uncertainty in stellar distances, but should not be over-interpreted at high-$|Z|$, especially in light of increasing distance uncertainties at high-$|Z|$ that can be on the order of or entirely exceed $s_2$.  We revisit this point in later subsections.

We also note that the spatial clustering of stars with $r_i \ne 1$ at marginal (1- or 2-$\sigma$) significance in Fig.~\ref{fig:yz_multipleScales8082} (a) and (b) may imply a higher significance for the structure itself.  That is, the high-$|Z|$ stars with 1- or 2-$\sigma$ significance are not merely scattered randomly about, but are clustered together into structures that far exceed $s_2$ in scale.  A closer look at this region in future data releases is warranted if distance uncertainties decrease.

Finally, we note that the seeming reversal in the overdensity/underdensity trend in Fig.~\ref{fig:yz_multipleScales8082} (c) and (d) is expected at larger values of $s_1$ and $s_2$.  Specifically, the distance between overdensities in panel (c) is about 0.5 kpc, so stars {\it between} the overdensities in panel (c) are thus most often paired with stars {\it in} the overdense regions of (c), resulting in the apparent inversion in panel (d).  In other words, stars in between the wave crests are positioned just right to pair with stars from {\it both} adjacent wave crests.

\subsection{Robustness Check: Reanalysis with Relative Parallax Error}

\begin{figure*}[]
  \begin{center}
    \subfloat[]{\includegraphics[scale=0.82]{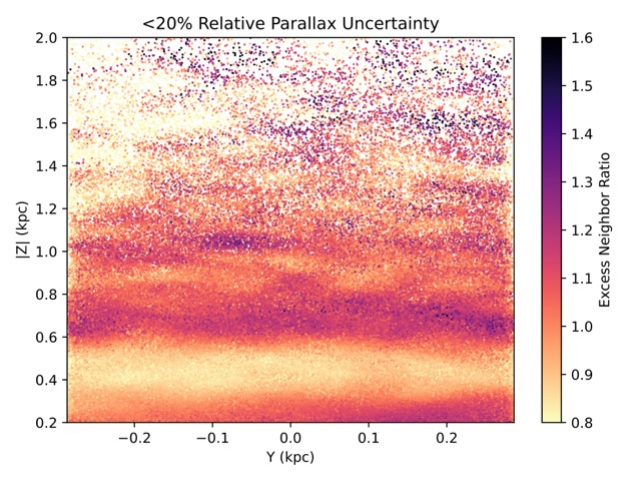}}
    \subfloat[]{\includegraphics[scale=0.82]{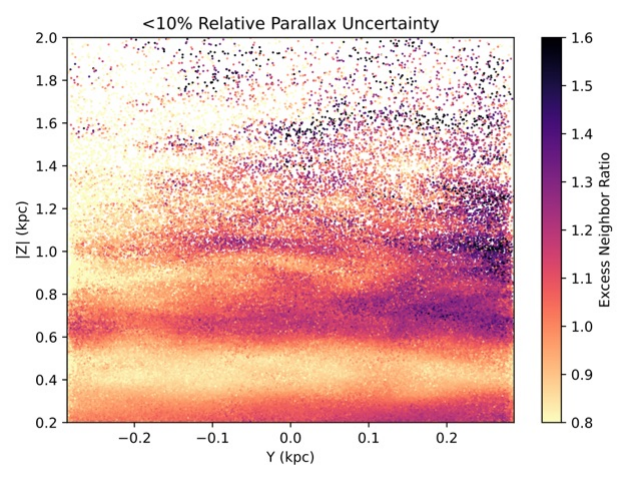}}
    \caption{
    A reanalysis of the region Fig.~\ref{fig:verticalWavesPOC}(b) for $s_1 = 10$ pc and $s_2 = 50$ pc, with additional relative parallax error cuts included.  Namely, we examine the effects of both
    (a) $\sfrac{\sigma_{\varpi}}{\varpi} < 0.20 $ and
    (b) $\sfrac{\sigma_{\varpi}}{\varpi} < 0.10$ parallax quality cuts.  However, including these additional requirements on the data imprints Gaia scan law patterns on the data.  Specifically, the more stringent quality cuts preferentially cut away stars in the $Y < 0$ region for the Northern stars
     while also preferentially cutting stars with $Y > 0$ away for the Southern stars.  This is driven by the number of observations (visibility periods used in the astrometric solution) for the stars in the North and South (see Fig.~\ref{fig:visperiods}) and is thus not indicative of true structure.  We thus opt to exclude parallax error quality cuts in line with \citet{hinkel2023two}.  
    }
  \label{fig:relprlxerr}
  \end{center}
\end{figure*}

\begin{figure*}[]
  \begin{center}
    \subfloat[]{\includegraphics[scale=0.82]{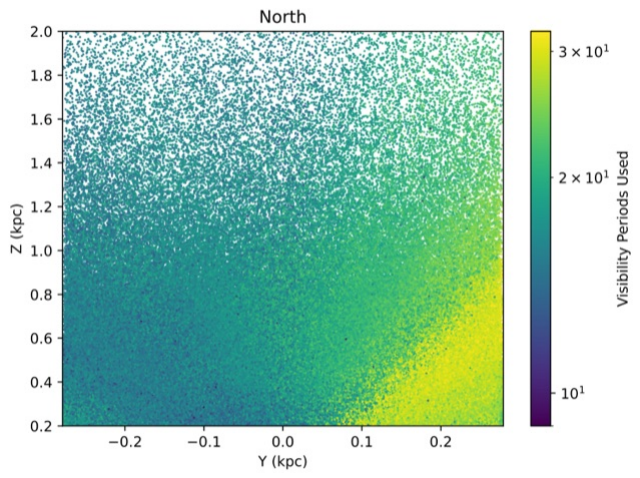}}
    \subfloat[]{\includegraphics[scale=0.82]{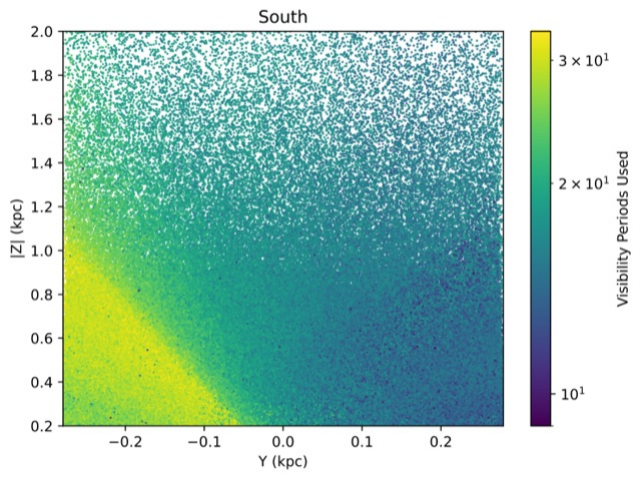}}
    \caption{
    The number of visibility periods used in the astrometric solution for the stars in the North (a) and South (b) drives the left/right effect in Fig.~\ref{fig:relprlxerr}.  We therefore do not employ parallax quality cuts and instead rely on the data selection method of \citet{GHY20} and \citet{HGY20}.
    }
  \label{fig:visperiods}
  \end{center}
\end{figure*}

In light of increasing distance uncertainties for high-$|Z|$ stars, it is worth examining if the structures we highlight in this manuscript persist with hard limits on the relative parallax error, $\sfrac{\sigma_{\varpi}}{\varpi}$.  In Fig.~\ref{fig:relprlxerr} panel (a) we keep only stars with $\sfrac{\sigma_{\varpi}}{\varpi} < 0.20 $, while in panel (b) we keep only stars with $\sfrac{\sigma_{\varpi}}{\varpi} < 0.10$.  The wave structure persists across both panels, but becomes more concentrated in the $Y > 0$ region due to Gaia scan law effects.  Namely, as we plot the positions of the Northern stars in our 2PCC visualizations, the fact that the spacecraft observed the $Y > 0$ region more often in the North (see Fig.~\ref{fig:visperiods} panel (a)) explains the larger number of neighbors in that region, as more stars satisfy the parallax quality cuts.  Moreover, in the South, the Gaia mission observed the stars more often in the $Y < 0$ region of our selection (see panel (b)), resulting in lower values of $r_i$ for that region, as this data contributed to the denominator of the Excess Neighbor Ratio, $r$.  In other words, the more often a star is observed, the more likely it is to have lower reported uncertainties and thus survive the parallax quality cuts, imprinting the scan law on the data.  While originally forgoing parallax error cuts to match the selection of \citet{hinkel2023two}, we reaffirm here that these cuts are not appropriate for our analysis.

\begin{figure*}[]
  \begin{center}
    \subfloat[]{\includegraphics[scale=0.82]{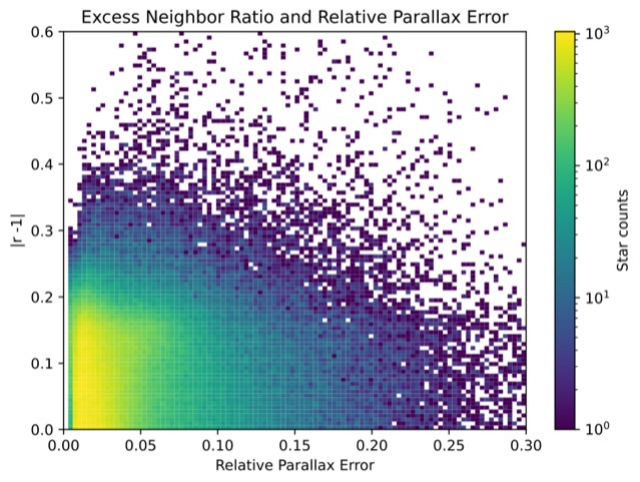}}
    \subfloat[]{\includegraphics[scale=0.82]{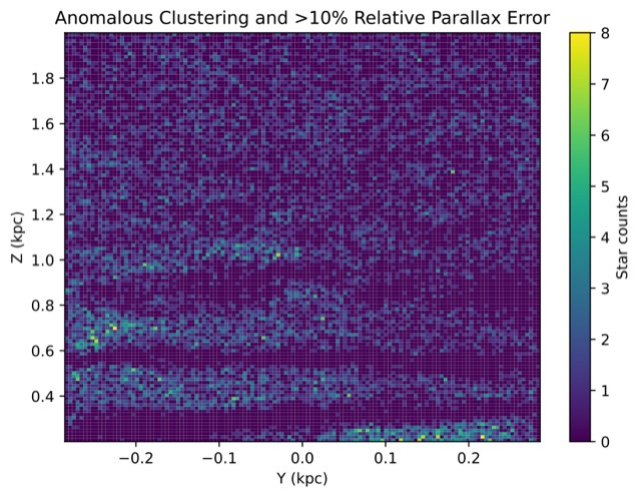}}
    
    \subfloat[]{\includegraphics[scale=0.82]{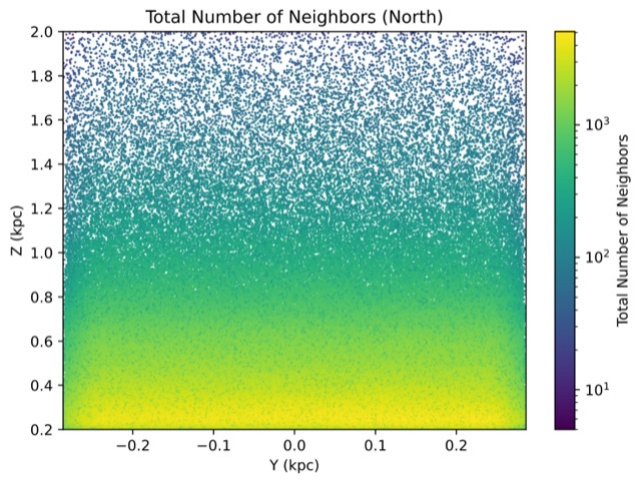}}
    \subfloat[]{\includegraphics[scale=0.82]{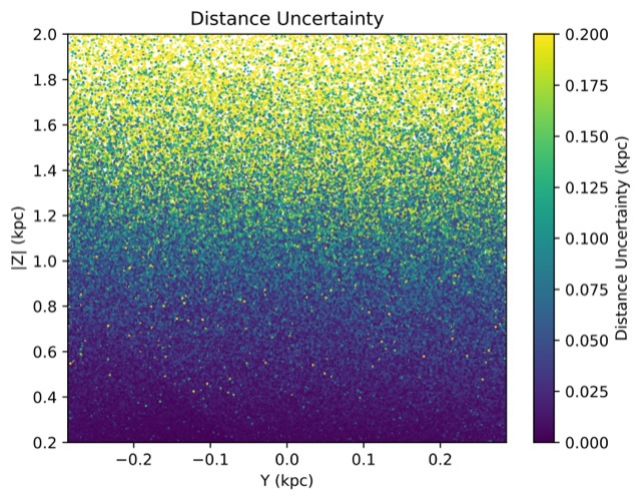}}
    \caption{
    For the region of Fig.~\ref{fig:verticalWavesPOC} with $s_1 = 10$ pc and $s_2 = 50$ pc, (a) anomalous values of $r_i$ are not skewed towards higher relative parallax error.  (b) Nonetheless, by isolating the stars with $|r_i - 1| > 0.1$ and a relative parallax error in excess of 10\%, we find only a small number of stars spread throughout our volume of interest.  (c) Compared to the total number of neighbors for each star in the North ($N_c$), these higher error points comprise only a small fraction of the data and cannot explain the structures we find.  (d) Nonetheless, both North and South stars show a clear trend of increasing distance uncertainty with $|Z|$, implying that no single choice of $s_1$ and $s_2$ can be used to uncover structure at all $|Z|$ simultaneously.  Additionally, no obvious clusters with anomalous uncertainties appear.
    }
  \label{fig:relprlxerr2}
  \end{center}
\end{figure*}

Still, by not cutting on the relative parallax error, we allow some stars to enter the analysis with larger uncertainties, so it is important to verify that the structures we find are not dominated by stars with large uncertainties.  To do so, we repeat the analysis for Fig.~\ref{fig:verticalWavesPOC}(b) and then look for correlations between the Excess Neighbor Ratio for each star and the relative uncertainty in the parallax for each star, which we include as Fig.~\ref{fig:relprlxerr2}(a).  Further, we select the stars that satisfy both $|1 - r| > 0.1$ and $\sfrac{\sigma_{\varpi}}{\varpi} > 0.10 $ and plot their positions in Fig.~\ref{fig:relprlxerr2}(b).  By comparing the total number of stars in the regions of the clumping that we observe (panel (c)) to these higher-error stars with anomalous excess neighbor ratios in panel (b), we conclude that the structures are not dominated by high error stars. For easier comparison with the distance scales we explore, we have also transformed the parallax uncertainties into the distance uncertainties in panel (d).

\subsection{A Note on Dust}

Additionally, it is important to note that extinction effects are unlikely to be the dominant driver of the structures we identify here. Indeed, by choking up on the data in both $R$ and $\phi$, the majority of our volume is above the densest dust mapped by \citet{green20193d}.  In particular, the over- and under-densities at $|Z| \sim 400$ and 700 pc lie along sight-lines with $|b| \gtrsim 49$ deg. and $|b| \gtrsim 63$ deg. respectively, placing the most significant structures in our analysis well above the bulk of the Galaxy's dust.  The lowest over-density in our analysis near $|Z| \sim 200$ pc does overlap with some of the regions of higher extinction mapped by \citet{green20193d}, so residual dust effects may contribute at the extreme low end of our sample.

\section{Results and Discussion} \label{sec:results}

We have introduced here a new tool for highlighting Galactic structure at specific length scales of interest.  Namely, we have described the processes behind the Two-Point Correlation Clustering Algorithm (2PCC) and have applied it to Gaia DR3 data.  In doing so,  
we have visualized the vertical waves \citep{widrow2012galactoseismology} in unprecedented detail, shown that substructure exists down to scales of order 10 parsecs, and revealed substantial azimuthal and radial variations in the vertical structure of the Milky Way.  

Critically, we {\it localize} the structure implied to exist via the 2PCF studies of \citet{hinkel2023two}.  Specifically, the first peak in each of the panels of Fig. 6 of \citet{hinkel2023two} corresponds to the tight clustering of stars within each crest of the vertical waves.  The first trough in those same panels corresponds to the pairings of stars in the under-dense regions with both of the adjacent, denser wave crests above and below it. The correlations at $z_{12} \approx 1.4$ kpc in Fig. 6 of \citet{hinkel2023two} seem to correspond to pairings made between the denser low-$Z$ wave crests and small clumps of stars at higher values of $Z$, though these higher-$Z$ clusters are not found to be highly significant in this work.  However, this higher-$Z$ structure qualitatively matches the findings of \citet{hinkel2023two} in their Fig. 8, exhibiting non-planar wave structure in contradistinction to the lower-$Z$ region.

In order to isolate the stars responsible for each of the peaks and troughs in \citet{hinkel2023two}, we have chosen various scales for $s_1$ and $s_2$ to explore the different structures. However, we caution against over-interpreting the high-$|Z|$ results given the larger parallax uncertainties in that region.  Although the smaller choices of $s_1$ and $s_2$ enabled a clear view of the microstructure of the low-$|Z|$ wave crests, these scales are smaller than the worst case distance uncertainties at high-$|Z|$.  

Nonetheless, we have explored the impact of different choices of $s_1$ and $s_2$ and parallax quality cuts and still find evidence of some marginally significant structure at higher-$|Z|$.  High-error stars indicated to be involved in a structure were shown to be only very small contributions to the structures we find.  
Examining the astrophysical characteristics of stars found to be in the wave structure may help to further understand its origins.
Revisiting the high-$|Z|$ region with Gaia DR4 may also prove interesting, as would investigating whether the azimuthal and radial variations we see are related to the phase spiral variations discussed by \citet{antoja2023phase} and \citet{hunt2022multiple}. 

Further, it would be interesting to explore if the less coherent structure we see at high-$|Z|$ could be consistent with the combination of small-scale and large-scale kicks explored by \citet{tremaine2023origin}.  In particular, their simulations produce phase-spiral structure that departs from the coherent spiral pattern at high-$|Z|$ (c.f. their Figure 4, e.g.).  As such, precise measurements of the velocities of the stars indicated to be involved in high-$|Z|$ structure in this work may help to better understand the phase-spiral.

The high-$|z|$ substructure depicted in Fig.~\ref{fig:verticalWavesPOC} is tantalizing, as it exists well away from the Galactic midplane, and no longer exhibits coherent vertical waves as in the disk region, though there are still some hints of planar, vertical waves.  It could be that stars perturbed to higher heights above the plane have larger vertical oscillations about the Galactic midplane and are thus more susceptible to disruption by halo substructure \citep[e.g.]{gilman2025dark, bland-hawthorne2016galaxy}, as they spend more time in this region.  If this is indeed the case, relating the vertical oscillation frequency of the participating stars to the time it takes for the halo substructure to meaningfully disrupt the waves could help to discriminate against Galaxy evolution models or even models of dark matter \citep[e.g.]{buckley2018gravitational, gardner2021milky}.  Further examination of this region with future 6-D phase space data could help to better understand the origins of these structures \citep{frankel2025iron} by first selecting stars indicated to belong to the structures in the more abundant 5-D astrometric data, and then joining any available 6-D data to those stars.  This avoids the issues of first requiring 6-D data availability and its associated selection effects.

Another possibility is that because the vertical waves still appear {\it somewhat} coherent in the radial and azimuthal directions, it could be that the azimuthal differences existed {\it before} the vertical wave was excited, or were perhaps induced during the same perturbation.  In this picture, a dynamically cold disk is perturbed, and any azimuthal variations in the perturbation excite different portions of the disk into differing vertical oscillations.  If this is indeed the case, we would expect each of the patterns to wrap up azimuthally with time due to the differential rotation of the disk \citep[e.g.]{bernet2025dynamics}, causing a tightly-banded vertical wave in the $X-Z$ and $Y-Z$ planes. This effect may be faintly visible in the higher-$|Z|$ region of Fig.~\ref{fig:verticalWavesPOC} and Fig.~\ref{fig:verticalWavesPOC_radial}, though it is of low significance when considered on a star-by-star basis.  Comparing population statistics for stars in the over-dense regions to those outside of the over-dense regions may help to better test this picture. 

Yet another intriguing possibility is that the vertical plane waves are superimposed on a breathing mode, perhaps similar to those proposed by \citet{carrillo2018milky}, \citet{ghosh2022age} or \citet{widmark2022mapping}.  In fact, Fig.~\ref{fig:verticalWavesPOC}(b) reveals azimuthal variations in the gradual undulation of each crest of the vertical wave, where the crest closest to the midplane is bent towards the plane (and out of our volume) for $Y \lesssim 0$, while the portion of this crest with $Y \gtrsim 0$ is bent away from the midplane.  The second crest ($|Z| \approx 0.7$ kpc) and the third crest ($|Z| \approx 1$ kpc) are slightly closer to one another for $Y \sim 0$ while they appear slightly further apart (though faintly!) for $Y \gtrsim 0$.  If different regions of the disk are subject to varied breathing motions, this could help explain the departure from planar waves in Fig.~\ref{fig:verticalWavesPOC}.  As the various wave crests are bent in different directions, we do not believe the effect is due solely to the warping of the Milky Way disk \citep[e.g.]{skowron2019three, chen2019intuitive}.

While tantalizing, these pictures all require further study and would benefit from velocity data and data on the astrophysical parameters of each star.  Fortunately, the 2PCC algorithm offers us a way to tag each star based on its participation in spatial structure at a distinct scale.  This ``dynamical tagging" of stars from {\it purely spatial data} could then allow for population studies between ``in-structure" stars and ``out-of-structure" stars.  In other words, could the stars identified with the statistical strength of the 3-D position data in Gaia DR3 then be joined with data for velocity, metallicity, and other critical information to understand if the structures are recent phenomena?  Do they move coherently?  Are they bluer or redder than the surrounding stars?  Could such information be useful in extending Gaia's radial velocity data set to new stars via machine learning approaches? Such an approach would be complementary to the Orbital Torus Imaging \citep{price2021orbital, horta2024orbital} and Element Abundance  \citep{widrow2025equilibrium} methods which require spectroscopic data from the start.

Indeed, the ``Great Wave" mapped by \citet{poggio2024great} is seen in stars thought to be very young (${\cal O}$(100) Myr) \citep{poggio2022chemical}, based on metallicities inferred from a robust XGBoost machine learning model \citet{andrae2023robust}.  
The radial wave mapped by \citet{yin2024wave} may also track younger stars, as color-magnitude cuts were made to select stars on the more luminous side of the main sequence (and thus possibly along younger isochrones) but were inadvertently omitted from the discussion in that paper\footnote{For completeness, the cuts in \citet{yin2024wave} were: $3.8557(G_{\rm BP} - G_{\rm RP}) - 0.4312 < M_G < 3.4251(G_{\rm BP} - G_{\rm RP}) + 1.517$ and $G < 18$, as opposed to $13 < G < 18$. These cuts are not employed in the  present work.}.  Indeed, the radial wave of \citet{yin2024wave} does not appear nearly as clearly when these cuts are not included, suggesting that the color-magnitude cuts select a distinct population of younger stars participating in the structure.  
It is conceivable, then, that spatial structure can first be isolated via a 2PCC analysis and later joined with data similar to that of \citet{andrae2023robust} so that the age of the structure can be estimated.  This may also offer a complementary and independent method of dating the perturbation that is thought to have triggered the corrugation waves in the Milky Way \citep{tepper2022galactic}.

\section*{Acknowledgments}
This material is based upon work supported by 
the NASA National Space Grant College and Fellowship
Program and the Kentucky Space Grant Consortium 
under NASA award number 80NSSC20M0047.

The authors thank Robert Riehemann for helpful discussions on the mathematical description of the algorithm, and thank the anonymous referees for constructive comments that have improved the presentation of this work.  A.H. also thanks Susan Gardner for helpful discussions regarding various aspects of this work.

A.H. thanks Pavadol Yamsiri and Joss Bland-Hawthorn for discussions from which the omitted cuts in \citet{yin2024wave} were noticed.

This work has made use of data from the European Space Agency (ESA) mission
{\it Gaia} (\url{https://www.cosmos.esa.int/gaia}), processed by the {\it Gaia}
Data Processing and Analysis Consortium (DPAC,
\url{https://www.cosmos.esa.int/web/gaia/dpac/consortium}). Funding for the DPAC
has been provided by national institutions, in particular the institutions
participating in the {\it Gaia} Multilateral Agreement.


\vspace{2\baselineskip}
\bibliography{mybib}

\enlargethispage{1.9cm}
\end{document}